\documentclass[prb,reprint,citeautoscript,superscriptaddress]{revtex4-1}
\usepackage{amsmath}
\usepackage{bm}
\usepackage{graphicx,color}
\usepackage{stix}
\usepackage[usenames,dvipsnames]{xcolor}

\newcommand{\vk}{\bm{k}}
\newcommand{\vT}{\bm{T}}
\newcommand{\imp}{\mathrm{imp}}
\newcommand{\bath}{\mathrm{bath}}

\definecolor{myblue}{rgb}{0,0,1}
\usepackage[breaklinks=true,colorlinks=true,linkcolor=myblue,urlcolor=myblue,citecolor=myblue]{hyperref}

\begin{document}
\title{Coupled-cluster impurity solvers for dynamical mean-field theory}
\author{Tianyu Zhu}
\affiliation{Division of Chemistry and Chemical Engineering, California Institute of Technology, Pasadena CA 91125}
\author{Carlos A. Jim\'{e}nez-Hoyos}
\affiliation{Department of Chemistry, Wesleyan University, Middletown CT 06457}
\author{James McClain}
\affiliation{Division of Chemistry and Chemical Engineering, California Institute of Technology, Pasadena CA 91125}
\author{Timothy C. Berkelbach}
\email{tim.berkelbach@gmail.com}
\affiliation{Department of Chemistry, Columbia University, New York NY 10027}
\affiliation{Center for Computational Quantum Physics, Flatiron Institute, New York NY 10010}
\author{Garnet Kin-Lic Chan}
\email{gkc1000@gmail.com}
\affiliation{Division of Chemistry and Chemical Engineering, California Institute of Technology, Pasadena CA 91125}

\begin{abstract}
  We describe the use of coupled-cluster theory as an impurity solver 
  in dynamical
mean-field theory (DMFT) and its cluster extensions.  We present numerical results
at the level of coupled-cluster theory with single and double excitations
(CCSD)
for the density of states and self-energies of cluster impurity
problems in the one- and two-dimensional Hubbard models.
Comparison to exact diagonalization
shows that CCSD produces accurate density of states and self-energies
at a variety of
values of $U/t$ and filling fractions.
However, the low cost allows for the use of many bath
sites, which we define by a discretization of the hybridization directly on
the real frequency axis.  We observe convergence of dynamical quantities using approximately 30 bath
sites per impurity site, with our largest 4-site cluster DMFT calculation using 120 bath sites. We
suggest coupled cluster impurity solvers will be attractive in ab initio
formulations of dynamical mean-field theory.
\end{abstract}

\maketitle

\section{Introduction}

Dynamical mean-field theory (DMFT)~\cite{Georges1992,Georges1996} and its
cluster extensions (such as cluster dynamical mean-field theory
(CDMFT)~\cite{Lichtenstein2000,Kotliar2001} and the dynamical cluster
approximation (DCA)~\cite{Hettler1998,Hettler2000}) approximate the
single-particle Green's function of an interacting quantum lattice Hamiltonian
using the self-energy of a self-consistent impurity model.  Computing the
impurity self-energy and Green's function is thus the main numerical task, and
falls to the so-called quantum impurity solver, the focus of this work.

In DMFT impurity models, the impurity sites retain the full interaction, while
the rest of the lattice is replaced by a self-consistent hybridization
$\mathbf{\Delta}(\omega)$.  Impurity solvers can be divided into two classes
based on how they treat this hybridization.  In the first class (diagrammatic),
which includes methods such as diagrammatic Monte Carlo (diagMC)~\cite{Kozik2010} and some forms
of continuous-time quantum Monte Carlo (CT-QMC)~\cite{Werner2006,Gull2008,Gull2011}, $\mathbf{\Delta}(\omega)$ is
directly included in the evaluation of the diagrams.  In the second class
(bath-based), which includes methods such as exact diagonalization (ED)~\cite{Caffarel1994,Capone2007},
configuration interaction (CI)~\cite{Zgid2011}, numerical and density matrix renormalization
group methods (NRG, DMRG)~\cite{Wilson1975,White1992,Bulla2008}, and the variational Gutzwiller ansatz~\cite{Gutzwiller1963}, the
hybridization is first unfolded into a discrete bath. The impurity and bath
sites then together define an impurity Hamiltonian, from which the Green's
function and self-energy can be computed from a finite system simulation.  At
zero temperature, this usually involves computing the impurity ground-state
wavefunction, and the impurity Green's function as a correlation function.  In
this work, we will explore an impurity solver based on coupled-cluster (CC)
theory~\cite{Cizek1966}, which falls into this second class of solvers.

The bottleneck in all bath-based methods is the number of sites in the impurity
Hamiltonian.  Even for a small number of impurity sites, one requires several
bath sites per impurity site to adequately discretize the hybridization. The
exponential complexity of exact diagonalization  limits calculations to about 16
sites in total, and thus to only small impurity clusters, typically with no more
than 4 impurity sites and 2 bath sites per impurity site. There are two common
strategies to ameliorate this bottleneck.  The first is to restrict the Hilbert
space in which the diagonalization is performed. This was explored by Zgid and
Chan with truncated CI~\cite{Zgid2011}, which defines a systematic selection of
a reduced set of Slater determinants in which to solve the quantum impurity
problem.  Bravyi~\cite{Bravyi2017} provided a formal justification for this truncation, showing
that for fixed impurity size, the Green's function can be converged by a linear
combination of Gaussian states with a cost that is polynomial in the accuracy.
CI solvers significantly extend the size of systems that can be treated by
diagonalization methods
particularly with respect to the number of bath sites~\cite{Zgid2012,lin2013efficient,lu2014efficient,Go2017}.
A second strategy is to parametrize the
impurity wavefunction through a non-linear ansatz.  NRG, DMRG, and the
variational Gutzwiller approximation adopt this latter approach.

The CC solvers we explore here also correspond to a non-linear
ansatz for the impurity wavefunction. CC theory has similar strengths to
configuration interaction, for example it can be used at zero-temperature as
well as finite-temperature~\cite{White2018} and with arbitrary interactions and couplings, but it
addresses several formal deficiencies of the truncated CI wavefunction. Most
importantly, the CC parametrization is size-extensive~\cite{Bartlett1981}. This means that (much
like tensor networks) one can represent product states of disjoint interacting
clusters, with a number of parameters linear in the number of clusters, instead
of the exponential number of parameters in exact diagonalization. When
interactions are not too strong, this makes the CC parametrization exponentially
more compact than the CI parametrization for a given accuracy. In ab initio
quantum chemistry calculations, CC theory is generally used in place of CI~\cite{Bartlett2007},
except when there are simultaneously strong interactions and a large number of
degenerate sites.

In this work, we study the performance of truncated CC theory as an impurity
solver in cluster dynamical mean-field theory calculations. We will use the 1D
and 2D Hubbard models, canonical models of correlated materials, as our test
systems.  We focus on the lowest order truncation in the CC theory of the
Green's function, namely, equation-of-motion CC theory with single and double
excitations (EOM-CCSD), a level of theory that is widely implemented and
available in quantum chemistry packages. To understand the quality of our
low-order CC truncation, we will compare  to dynamical quantities obtained from an exact
diagonalization solver with small bath discretizations. We will further exploit the
low cost of the CC solver to carry out calculations with a large number of bath sites. 
Overall, we seek to shed light on the regimes in which low-order CC methods are a
promising class of impurity solvers for dynamical mean-field theories.
After this work was submitted, Shee and Zgid have presented related work~\cite{Shee2019} that also explores the use of CC impurity solvers in Green's function embedding methods. 

\section{Theory Recapitulation}

\subsection{Cluster Dynamical Mean-Field Theory}
\label{ssec:cdmft}

Cluster dynamical mean-field theory (CDMFT) has been extensively
reviewed~\cite{Maier2005,Kotliar2006,Held2007} and we present only a minimal description sufficient for
the numerical considerations in our work. We first consider a general
translationally invariant lattice Hamiltonian with one- and two-particle
interactions
\begin{align}
\hat{H} = \sum_{pq} h_{pq} a^\dag_p a_q 
    + \frac{1}{2} \sum_{pqrs} V_{pqrs} a^\dag_p a^\dag_q a_r a_s
\end{align}
where $p,q,r,s$ label lattice sites (including spin) and $a^{(\dag)}$ are fermion annihilation 
(creation) operators. We then consider a computational unit cell
$\mathcal{C}$ with $N_\mathcal{C}$ cluster sites, that tiles the lattice through
a set of translation vectors $\vT$. Taking $\vk$ as a corresponding reciprocal
space vector taken from an evenly spaced mesh of $N_k$ points in the first Brillouin zone of
$\mathcal{C}$, the reciprocal space Hamiltonian becomes
\begin{equation}
\begin{split}
  \hat{H} &= \sum_{pq \in \mathcal{C}} \sum_{\vk}  \tilde{h}_{pq}(\vk) \tilde{a}^\dag_{p\vk} \tilde{a}_{q\vk} \\
  &+ \frac{1}{2} \sum_{pqrs \in \mathcal{C}} \sum_{\vk_p\vk_q\vk_r} 
        \tilde{V}_{pqrs}(\vk_p,\vk_q,\vk_r) 
        \tilde{a}^\dag_{p\vk_p} \tilde{a}^\dag_{q\vk_q} \tilde{a}_{r\vk_r} \tilde{a}_{s\vk_p+\vk_q-\vk_r}
\end{split}
\end{equation}
with $\tilde{a}^{(\dag)}_{p\vk} = \frac{1}{\sqrt{N}_k} \sum_{\vT} a^{(\dag)}_p e^{i \vk\cdot\vT}$ 
where $\tilde{h}$ and $\tilde{V}$ are the matrix elements of $h$ and $V$ in the
symmetry-adapted basis. The non-interacting and interacting Green's functions,
$\mathbf{g}(\vk,\omega)$ and $\mathbf{G}(\vk,\omega)$, are block diagonal in
reciprocal space, and are related by the block-diagonal lattice self-energy
$\mathbf{\Sigma}(\vk,\omega)$ via the Dyson equation
\begin{align}
\mathbf{G}(\vk,\omega) = \left[(\omega+\mu)\mathbf{1}-\mathbf{\tilde{h}}(\vk)
    -\mathbf{\Sigma}(\vk,\omega)\right]^{-1}.
\end{align}
The cellular Green's function is related to the reciprocal space Green's
function by
\begin{align}
\mathbf{G}_\mathcal{C}(\omega) = N_k^{-1} \sum_{\vk} \mathbf{G}(\vk,\omega).
\end{align}

The key quantity to approximate is the lattice self-energy that contains the
effects of interactions.  In CDMFT, the lattice self-energy is taken to be equal
to the self-energy of an impurity with $N_\mathcal{C}$ sites, i.e.
\begin{align}
  \mathbf{\Sigma}(\vk,\omega) = \mathbf{\Sigma}_\imp(\omega) \label{eq:cdmft_sigma}
\end{align}
The impurity model is characterized by a cellular hybridization
$\mathbf{\Delta}(\omega)$ that describes the delocalization effects between the
cell and the lattice. Defining the cellular non-interacting Hamiltonian
$\hat{h}_\mathcal{C}$ as
\begin{align}
  \hat{h}_\mathcal{C} &= \sum_{pq\in \mathcal{C}} h_{pq} a^\dag_p a_q,
  \end{align}
the hybridization follows as
\begin{align}
  \mathbf{\Delta}(\omega) = (\omega+\mu)\mathbf{1}-\mathbf{h}_\mathcal{C} -\mathbf{\Sigma}_\mathrm{imp}(\omega) - \mathbf{G}^{-1}_\mathcal{C}(\omega)
\label{eq:hybridization}
\end{align}
The impurity Green's function $\mathbf{G}_\imp(\omega)$ is formally defined from
the zero-temperature generating functional $W[\mathbf{J}]$
\begin{align}
  W &= \int \! \! \! \int \mathcal{D}\mathbf{c} \mathcal{D}\bar{\mathbf{c}} \ e^{i S(\mathbf{J})} \\
\begin{split}
  S &= \int \! \! \! \int dt dt'  \left[ \bar{\mathbf{c}}^T(t) [(i\partial_t - \mathbf{h}_\mathcal{C})  \delta(t-t')- \mathbf{\Delta}(t,t')\right. \\
&\hspace{1em} \left.    + \mathbf{J}(t,t')] \mathbf{c}(t')\right] + V[\mathbf{c},\bar{\mathbf{c}}]
\end{split}
\end{align}
where $\mathbf{c}, \bar{\mathbf{c}}$ are
vectors of $N_\mathcal{C}$ Grassmann variables, 
$V[\mathbf{c},\bar{\mathbf{c}}]$ is the interaction contribution to the action 
$S$, 
 $\mathbf{G}_\imp(t,t') = \delta W/\delta \mathbf{J}(t,t')$, and 
$\mathbf{G}_\imp(\omega) = \frac{1}{2\pi} \int dt \ \mathbf{G}_\imp(0,t) e^{i\omega t}$.

From the impurity Green's function, $\mathbf{\Sigma}_\imp(\omega)$ follows as
\begin{align}
\mathbf{\Sigma}_\imp(\omega) 
    =\left[(\omega+\mu)\mathbf{1}-\mathbf{h}_\mathcal{C} 
        - \mathbf{\Delta}(\omega)\right] -\mathbf{G}_\imp^{-1}(\omega).
\label{eq:sigmaimp}
\end{align}
Using $\mathbf{\Sigma}_\imp(\omega)$ as the lattice self-energy in
\eqref{eq:cdmft_sigma} leads to a new lattice Green's function and
hybridization, and thus a new impurity Green's function. The self-consistency in
CDMFT is then achieved when the cellular Green's function and impurity Green's
function agree, 
\begin{align}
  \mathbf{G}_\imp(\omega) = \mathbf{G}_\mathcal{C}(\omega)
  \end{align}
Note that after self-consistency, if the primitive cell of the lattice is smaller than the computational cell, the cellular Green's function may
break translational invariance. There are variety of related formulations~\cite{Biroli2004,Capone2004}, that restore the translational invariance (including the dynamical cluster
approximation~\cite{Hettler1998,Hettler2000}) but we will not use them here.

In the bath-based CDMFT, the hybridization function is represented by a set of
fictitious bath sites and couplings.  Formally, we consider each element
of $\mathbf{\Delta}(\omega)$ as the Hilbert transform~\cite{DeVega2015}
\begin{align}
\mathbf{\Delta}(\omega) = \int \! d\varepsilon \ 
    \frac{\mathbf{J}(\varepsilon)}{\omega -\varepsilon} \label{eq:hilbert}
\end{align}
with the spectral density
\begin{equation}
\mathbf{J}(\omega) = -\frac{1}{\pi} \mathrm{Im}\mathbf{\Delta}(\omega+i\eta). 
\label{eq:Jw}
\end{equation}  
A bath parametrization can be considered a discrete representation
of the above integral, where $\varepsilon$ and $\mathbf{J}(\varepsilon)$ take
on a discrete set of values given by the bath energies and couplings.
One common choice is
to approximate $\mathbf{\Delta}(\omega)$ along the imaginary axis where it is
smooth and can be more easily fit to a bath representation by numerical optimization of
a cost function~\cite{Caffarel1994}. This is beneficial
for exact diagonalization (and related) solvers which can only handle a very
small number of bath sites but can lead to a loss of accuracy when analytically
continuing to the real axis for spectral computation~\cite{Liebsch2012,Koch2008,Zgid2011,wolf2015imaginary}.
Another choice is to
approximate $\mathbf{\Delta}(\omega)$ along the real axis directly, which is
commonly done with NRG and DMRG solvers~\cite{Bulla2008,Peters2011,Ganahl2014,Ganahl2015}.  
Here we view the Hilbert transform as
a quadrature along the real axis 
\begin{align}
\mathbf{\Delta}(\omega) = \sum_{n=1}^{N_\omega} w_n 
    \frac{\mathbf{J}(\varepsilon_n)}{\omega -\varepsilon_n} 
\end{align}
where $w_n$ are quadrature weights for $N_\omega$ integration grid points. Then, we can define 
an approximate hybridization of the form
\begin{align}
\label{eq:hybridization_disc}
{\tilde{\Delta}_{pq}}(\omega)
    = \sum_{n=1}^{N_\omega} \sum_{k=1}^{N_\mathcal{C}} 
        \frac{V_{p,k}^{(n)} V_{q,k}^{(n)}}{\omega-\varepsilon_{n}}
\end{align}
where $\mathbf{V}^{(n)} = [w_n \mathbf{J}(\varepsilon_n)]^{1/2}$.
This leads to the discrete impurity Hamiltonian,
\begin{equation}
\begin{split}
\hat{H}_\imp &= \hat{H}_\mathcal{C} \\
&+ \sum_{n=1}^{N_\omega} \sum_{k=1}^{N_\mathcal{C}} \left[ 
    \varepsilon_n a_{nk}^\dagger a_{nk} 
    + \sum_p \left(V_{p,k}^{(n)} a_p^\dagger a_{nk} + \mathrm{H.c.}\right)\right]
\end{split}
\end{equation}
where $\hat{H}_\mathcal{C}$ is the cellular Hamiltonian with interactions and
the creation and annihilation operators indexed by $nk$ in the last two terms act on the
fictitious bath space.  
From the impurity Hamiltonian, the Green's functions can then be
defined as correlation functions.  For example, the addition and removal parts
of the impurity Green's function $\mathbf{G}_\imp (\omega)$ are given by 
\begin{subequations}
\begin{align}
 G^+_{pq}(\omega) &=  \langle \Psi_0 | {a}_p  
    \left[\omega+\mu - (\hat{H}_\imp-E)+i\eta\right]^{-1}  {a}^\dag_q | \Psi_0\rangle \\
  G^-_{pq}(\omega) &= \langle \Psi_0 | {a}^\dag_q  
    \left[\omega+\mu - (E-\hat{H}_\imp)+i\eta\right]^{-1}  {a}_p | \Psi_0\rangle
\end{align}
\label{eq:exactgf}
\end{subequations}
The goal of the impurity solver in the bath-based representation is thus to
approximate the impurity model ground-state wavefunction $|\Psi_0\rangle$ and
the impurity Green's function $\mathbf{G}_\imp (\omega)$ via the expressions in \eqref{eq:exactgf}. We next describe
how this is done in CC theory.
 
\subsection{Coupled-cluster theory of the ground-state and Green's function}

\label{sec:cc}

We now describe the basics of CC theory. Detailed discussions of ground-state CC
and the CC Green's function can be found in Refs.~\onlinecite{Shavitt2009,Scuseria1988,Nooijen1992,Nooijen1993,McClain2016,Bhaskaran-Nair2016}. The CC ansatz is
an exponential parametrization of the wavefunction,
\begin{align}
  |\Psi_0\rangle = e^{\hat{T}} |\Phi_0\rangle
\end{align}
where $|\Phi_0\rangle$ is a reference determinant, and $\hat{T}$ is the cluster excitation operator
\begin{align}
  \hat{T} &= \sum_{ia} t_{i}^{a} a^\dag_{a} a_i + \frac{1}{4} \sum_{ijab} t_{ij}^{ab} a^\dag_a a^\dag_b a_i a_j + \ldots \notag\\
   &= \hat{T}_1 + \hat{T}_2 + \ldots \label{eq:gcc}
\end{align}
where indices $i,j,\ldots$ and $a,b, \ldots$, label particle (p) and hole (h)
orbitals in the reference determinant, and $t_{i}^{a}, t_{ij}^{ab}, \ldots$ are
1p1h, 2p2h, etc.~cluster amplitudes. 
Truncating the CC amplitudes at 1p1h, 2p2h, etc.~gives the coupled-cluster
singles (CCS), coupled-cluster singles and doubles (CCSD), etc. approximations.
CCS constitutes a mean-field ansatz.  Thus we use the CCSD 
ansatz, the lowest truncation that includes correlations, in this work.

Given a Hamiltonian $\hat{H}$, the reference determinant $|\Phi_0\rangle$ is
often chosen to be the mean-field  (Hartree-Fock) ground-state of $\hat{H}$
(although including $e^{\hat{T}_1}$ in the cluster operator renders the
approximation somewhat insensitive to the choice of determinant, as an arbitrary
determinant satisfies $|\Phi'\rangle = e^{\hat{T}_1}|\Phi_0\rangle$). 
The amplitudes are chosen to satisfy the projected CC
equations; for CCSD, these are
\begin{align}
  E &= \langle \Phi_0 |e^{-\hat{T}} \hat{H} e^{\hat{T}}|\Phi_0\rangle \notag\\
   0 &= \langle \Phi_{i}^a |e^{-\hat{T}} \hat{H} e^{\hat{T}}|\Phi_0\rangle \notag\\
   0 &= \langle \Phi_{ij}^{ab} |e^{-\hat{T}} \hat{H} e^{\hat{T}}|\Phi_0\rangle 
 \end{align}
where $\langle \Phi_{i}^a| = \langle \Phi_0|a_ia^\dag_a \ldots$ and $E$ is the
(non-variational) CC approximation to the ground-state energy. The operator
$e^{\hat{T}}$ defines a similarity transformation.  Thus $\bar{H}=e^{-\hat{T}}
\hat{H} e^{\hat{T}}$ is an effective Hamiltonian whose mean-field energy is $E$,
or equivalently, the CC wavefunction is a product state of similarity
transformed quasi-particles, defined by the quasi-particle operators
$\bar{a}^\dag_i = e^{-\hat{T}} a^\dag_i e^{\hat{T}}$.

The advantages of truncated CC versus truncated CI derive from the exponentiation of the cluster operator. A CI wavefunction can be
written in similar notation as
\begin{align}
|\Psi_0\rangle &= \hat{C} |\Phi_0\rangle \notag\\
\hat{C}&    = 1+  \sum_{ia} c_{i}^{a} a^\dag_a a_i + \frac{1}{4} \sum_{ijab} c_{ij}^{ab} a^\dag_a a^\dag_b a_i a_j + \ldots \notag\\
&= 1+ \hat{C}_1 + \hat{C}_2 + \ldots
\end{align}
The truncated $\hat{C}$ can be seen as a linearization of a truncated $e^{\hat{T}}$,
or equivalently, the terms in $\hat{T}$ are the cumulants of $\hat{C}$, rewriting the latter as products of cluster excitations, e.g.
$\hat{C}_2 = \frac{1}{2}\hat{T}_1^2 + \hat{T}_2$.
The extensivity of the truncated CC approximation also derives from the exponential structure, since for a system $AB$ consisting of two separated
clusters of sites $A$ and $B$, we can write $e^{\hat{T}_{AB}} | \Phi_{AB}\rangle = e^{\hat{T}_A}|\Phi_A\rangle e^{\hat{T}_B}|\Phi_B\rangle$.

If the interactions between particle-hole excitations are not too strong, we
expect higher-order cluster operators to become small. In this situation, the
truncated CC approximation provides a significant improvement over the truncated
CI ansatz. Low-order CC truncations are accurate if the state of interest can be
described in terms of low-order products of fluctuations around the chosen
reference mean-field state $|\Phi_0\rangle$. 
In this work, we will consider two possible CCSD formulations based on
different reference mean-field states corresponding to restricted 
Hartree-Fock (RHF) and unrestricted Hartree-Fock (UHF).  
In restricted CCSD (RCCSD), the reference mean-field state is
required to be a singlet ($S=0$) eigenfunction of $\hat{S}^2$, and
the cluster operators $\hat{T}_1, \hat{T}_2$ each commute with $\hat{S}^2$.
In unrestricted CCSD (UCCSD), the reference mean-field state is only required to be an eigenfunction of $\hat{S}_z$,
and the cluster operators commute with $\hat{S}_z$. Consequently, spin symmetry can be
broken in UCCSD, for example, to yield an antiferromagnetic state.
RCCSD is expected to be most accurate when the interacting state can be described by
fluctuations around a paramagnetic state, while UCCSD
is more appropriate to describe fluctuations around an (anti)ferromagnetic state.

From the CC ground state, we then compute the Green's function as a correlation
function.  Using the cluster operator, similarity transformed Hamiltonian, and
quasiparticle operators, we can rewrite the exact expressions for the Green's
function in \eqref{eq:exactgf} in the equation-of-motion coupled-cluster
(EOM-CC) form~\cite{Stanton1993,Monkhorst2009,Krylov2008}
\begin{subequations}
\begin{align}
\begin{split}
G^+_{pq}(\omega) &= \sum_{mn} \langle \Lambda_0 | \bar{a}_p |\Phi_m\rangle 
\langle \Phi_n| \bar{a}^\dag_q |  \Phi_0 \rangle \\
&\hspace{1em}\times
    \langle \Phi_m| 
        \left[\omega +\mu - (\bar{H}_\imp-E)+i\eta\right]^{-1} |\Phi_n\rangle 
\end{split}
\\
\begin{split}
G^-_{pq}(\omega) &= \sum_{mn} \langle \Lambda_0 | \bar{a}^\dag_q |\Phi_m\rangle 
    \langle \Phi_n| \bar{a}_p | \Phi_0\rangle \\
&\hspace{1em}\times
    \langle \Phi_m | 
        \left[\omega +\mu - (E-\bar{H}_\imp)+i\eta \right]^{-1} |\Phi_n \rangle 
\end{split}
\end{align}
\end{subequations}
where $\langle\Lambda_0|$ is the left eigenstate of $\bar{H}$, and $\sum_m
|\Phi_m\rangle \langle \Phi_m| = 1$, where $|\Phi_m\rangle$ is a determinant.
Defining response vectors $|R^\pm_p(\omega)\rangle$,
\begin{subequations}
\label{eq:ccgf_resp}
\begin{align}
|R^+_q(\omega)\rangle &= \hat{P}^+ 
    \left[\omega + \mu - (\bar{H}_\imp-E)+i\eta\right]^{-1}
    \hat{P}^+ \bar{a}^\dagger_q | \Phi_0\rangle \\
|R^-_p(\omega)\rangle &= \hat{P}^- 
    \left[\omega + \mu - (E-\bar{H}_\imp)+i\eta\right]^{-1}
    \hat{P}^- \bar{a}_p | \Phi_0\rangle,
\end{align}
\end{subequations}
where $\hat{P}^\pm$ is the projector onto 1p, 2p1h, ...~states or 1h, 2h1p,
...~states, allows the CC Green's functions to be efficiently computed as
\begin{subequations}
\label{eq:ccgf}
\begin{align}
G^+_{pq}(\omega) &= \langle \Phi_0|\bar{a}_p|R^+_q(\omega)\rangle \\ 
G^-_{pq}(\omega) &= \langle \Phi_0|\bar{a}^\dag_q|R^-_p(\omega)\rangle. 
\end{align}
\end{subequations}

A truncated EOM-CC approximation to the
Green's function contains two truncations, one of the $\hat{T}$ operator that
defines the ground-state, and another of the resolution of the identity
determinants in the Lehmann sum. In this work we will use the EOM-CCSD
approximation, where $\hat{T}$ corresponds to the ground-state CCSD truncation,
and where $|\Phi_m\rangle$, $|\Phi_n\rangle$ are restricted to 1h, 2h1p
($\mathbf{G}^-$) and 1p, 2p1h ($\mathbf{G}^+$) excitations out of the reference
determinant. Note that
\begin{subequations}
\begin{align}
  \bar{a}_p |\Phi_0\rangle &= (a_p + [a_p, \hat{T}]) | \Phi_0\rangle \\
  \bar{a}^\dag_p |\Phi_0\rangle &= (a^\dag_p + [a^\dag_p, \hat{T}]) | \Phi_0\rangle 
\end{align}
\end{subequations}
thus the truncation of $\hat{T}$ at the CCSD level implies that
$\bar{a}_p|\Phi_0\rangle$ and $\bar{a}^\dag_p|\Phi_0\rangle$ are expressible in
terms of 1h, 2h1p and 1p, 2p1h spaces, respectively.

From the Green's function one can compute all one-particle expectation values
(such as the particle number and the single-particle density matrix) as well as
the total energy, from the Migdal formula~\cite{Galitskii1958} 
$E = -\frac{1}{2\pi}\mathrm{Im}\int_{-\infty}^{\mu} d\omega \mathrm{Tr}
(\omega\mathbf{1} + \mathbf{h}) \mathbf{G}(\omega)$. 
However, EOM-CCSD is not a conserving approximation~\cite{Lange2018}, thus the energy computed
using the Migdal formula is different from the ground-state CC energy $E$.
Nonetheless, the single-particle density matrix $\gamma_{pq} = (2\pi i)^{-1}
\int_{-\infty}^{\mu} (G^-_{pq}(\omega) + G^+_{pq}(\omega))$ is correctly
normalized and equal to its definition as an energy derivative of the CC energy
functional.
Furthermore, the EOM-CCSD Green's function is not strictly causal.
One implication of this is that the impurity self-energy calculated via 
Eq.~(\ref{eq:sigmaimp})
may not have an imaginary part that is negative definite, similar to the
behavior observed in adaptively truncated CI solvers~\cite{Go2017}. A futher
discussion of this point and partial solution is described in the Appendix.

The computational cost of ground-state CCSD for a general (e.g.~quantum
chemistry) two-particle interaction is $\mathcal{O}(N^6)$ where $N$ is the
number of sites (more specifically, the cost is $\mathcal{O}(o^3v^3+o^2v^4)$
where $o$ is the number of electrons and $v=N-o$ is the number of unoccupied
states).  There are some special
considerations, however, when using CC to determine the ground-state of the
CDMFT impurity Hamiltonian.  The impurity Hamiltonian only has two-particle
interactions on the impurity sites when working in the site basis.
However, the CC equations are usually implemented in the mean-field molecular
orbital basis, for which the impurity model has two-particle interactions over
all orbitals. Although we do not take advantage of it here, in principle, using
the locality of the interaction in the site basis can significantly lower the computational cost of CCSD,
particularly if we consider the scaling of
the cost when increasing the number
of bath sites while the number of impurity sites is kept fixed.
Another consideration is that the CC
theory presented here is a zero-temperature pure state theory, with a fixed
particle number. This is in contrast to the grand canonical formulation of
(C)DMFT in Sec.~\ref{ssec:cdmft}. This means that it is necessary to search
over all particle numbers (and spin sectors) in the ground-state calculations
of the impurity to find the quantum numbers of the (lowest energy) ground state,
as done with other zero-temperature impurity solvers~\cite{Zgid2011}.
  
The cost of EOM-CCSD (without accounting for locality in the interaction) is 
$\mathcal{O}(N^5)$ per response vector and thus
$\mathcal{O}(N_\omega N_\mathcal{C} N^5)$ for all elements of the frequency-dependent
impurity Green's function. Computing the response vector can be done either by
solving a system of linear equations or by computing $|R^\pm_p(t)\rangle$ in the
time domain followed by Fourier transformation (similarly as done in td-DMRG
solvers~\cite{White2004,ronca2017time}). We have found that a generalized minimum residual (GMRES) solver~\cite{Saad1986} or a
simplified two-parameter generalized conjugate residual method with inner
orthogonalization and outer truncation (GCROT($m,k$))~\cite{DeSturler1999} works well for the linear
equations and converges in $\mathcal{O}(10)$ iterations when the response vector
at a nearby frequency is used to initialize the solution for the response vector
at a new frequency and the matrix diagonal is used as a preconditioner. 

Once the impurity Green's function is computed, it may then be used in all the
expressions in Sec.~\ref{sec:cc}. We note that
the paramagnetic formulation of CDMFT  is incompatible
with symmetry breaking that may occur in the UCCSD impurity solver. To
study a paramagnetic phase using the UCCSD solver, we spin-average the 
impurity Green's function, $\mathbf{G}_\mathrm{imp} = \sum_{\sigma = \uparrow,\downarrow}  G^{\sigma\sigma}_\mathrm{imp}$.

\subsection{Algorithm}

For completeness, we outline the full computational procedure for our CDMFT
algorithm at fixed chemical potential using a CC solver. The loop is
initialized with the Hartree-Fock hybridization. 
\begin{itemize}
\item[1.] Discretize hybridization via \eqref{eq:hybridization_disc_sym}.
\item[2.] Solve impurity problem with Hartree-Fock at fixed $\mu$ (electron
number can change).
\item[3.] Calculate impurity Green's function with CC via \eqref{eq:ccgf_resp}
and \eqref{eq:ccgf} using GMRES or GCROT($m,k$).
\item[4.] Calculate hybridization $\mathbf{\Delta}(\omega)$ via
  \eqref{eq:hybridization} and check for convergence. 
If not converged, update hybridization (using DIIS~\cite{Pulay1980}) and return to 1 .
\end{itemize}
If results are desired at a fixed occupancy, then the occupancy
can be calculated via $n = \mathrm{Tr}\bm{\gamma}_\imp/N_\mathcal{C}$ using the 
ground-state CC solution.  If the occupancy does not equal the target
occupancy, then the chemical potential is updated and the CDMFT loop
is repeated.

We have implemented the above algorithm using
the HF, CCSD, and EOM-CCSD routines from the PySCF quantum chemistry 
package~\cite{Sun2018}.

\section{Applications to the 1D and 2D Hubbard models}

\label{sec:hubbard}
The Hubbard model~\cite{Hubbard1963} is defined by the lattice Hamiltonian
\begin{align}
\hat{H} = -t\sum_{\langle pq\rangle,\sigma} a^\dag_{p\sigma} a_{q\sigma} 
    + U \sum_p n_{p\uparrow} n_{p\downarrow}
\end{align}
where $n_{p\sigma} = a^\dag_{p\sigma} a_{p\sigma}$. 
It is the canonical model for DMFT
studies as there are no nonlocal interactions.

\subsection{Implementation in the 1D and 2D Hubbard models}
\label{ssec:implementation}

We will consider 1 and 2 site clusters for the 1D Hubbard model and 1 and 4 site (2$\times$2) 
clusters for the 2D Hubbard model.
The 2 site and 4 site clusters contain
additional point group symmetry which allows us to define a symmetry-adapted
impurity orbital basis, associated with operators 
\begin{subequations}
\begin{align}
a^\dag_{\Gamma_+} &= \frac{1}{\sqrt{2}}(a^\dag_1 + a^\dag_2) \\
a^\dag_{\Gamma_-} &= \frac{1}{\sqrt{2}}(a^\dag_1 - a^\dag_2)
\end{align}
\end{subequations}
for two sites,
and 
\begin{subequations}
\begin{align}
a^\dag_{\Gamma_1} &= \frac{1}{2}(a^\dag_1 + a^\dag_2 + a^\dag_3 + a^\dag_4) \\  
a^\dag_{\Gamma_2} &= \frac{1}{2}(a^\dag_1 + a^\dag_2 - a^\dag_3 + a^\dag_4) \\
a^\dag_{\Gamma_3} &= \frac{1}{2}(a^\dag_1 - a^\dag_2 - a^\dag_3 + a^\dag_4) \\
a^\dag_{\Gamma_4} &= \frac{1}{2}(a^\dag_1 - a^\dag_2 + a^\dag_3 - a^\dag_4)
\end{align}
\end{subequations}
for four sites~\cite{Liebsch2008,Liebsch2009}.
The impurity Green's function, self-energy, and hybridization are diagonal in
the symmetry-adapted orbital basis.
The diagonal hybridization leads to a simpler form of
the bath discretization~\eqref{eq:hybridization_disc}, which can now be written
as
\begin{align}
\label{eq:hybridization_disc_sym}
\tilde{\Delta}_\Gamma(\omega)
    = \sum_{n=1}^{N_\omega} \sum_{k=1}^{N_\mathcal{C}} 
        \frac{|V_{\Gamma,nk}|^2}{\omega-\varepsilon_{n}}.
\end{align}
The one-electron couplings in the site basis are
$V_{p,nk} = \sum_{\Gamma} U_{p,\Gamma} V_{\Gamma,nk}$
where $\mathbf{U}$ is the frequency-independent matrix of symmetry-adapted eigenvectors associated
with the change of basis.

The bath energies and weights used to discretize the hybridization are
chosen according to Gauss-Legendre quadrature on the interval $[-7t+U/2,
+7t+U/2]$ for the 1D Hubbard model and $[-9t+U/2,+9t+U/2]$ for the 2D Hubbard
model.  In almost all calculations, we achieve convergence using a small
imaginary broadening $\eta/t = 0.1$.  In some strong-coupling cases where
convergence was especially challenging, we use a slightly large value $\eta/t =
0.2$.  This latter value results in a weaker hybridization and also facilitates
the solution of the CC linear response equations~\eqref{eq:ccgf_resp}, similar
to previous observations with DMRG solvers~\cite{Peters2011}. At convergence,
quantities on the real frequency axis are plotted with a larger broadening
$\eta/t = 0.5$, for visual clarity only.

\subsection{1D Hubbard model}

We first consider the 1D Hubbard model. 
We present results at $U/t=2$ and $U/t=6$ (weak and strong coupling relative to the single-particle bandwidth of $4t$).
We show the impurity density of states 
(DOS) on the real frequency axis $\rho(\omega) = -(\pi N_\mathcal{C})^{-1} \mathrm{Tr} \mathrm{Im} \mathbf{G}(\omega)$
and the imaginary part of the impurity
self-energy on the imaginary frequency axis $\mathrm{Im}\Sigma_{11}(i\omega_n)$.
For visual clarity, the DOS is plotted with a broadening of $\eta/t=0.5$.

\begin{figure}[t]
\centering
\includegraphics[scale=1.0]{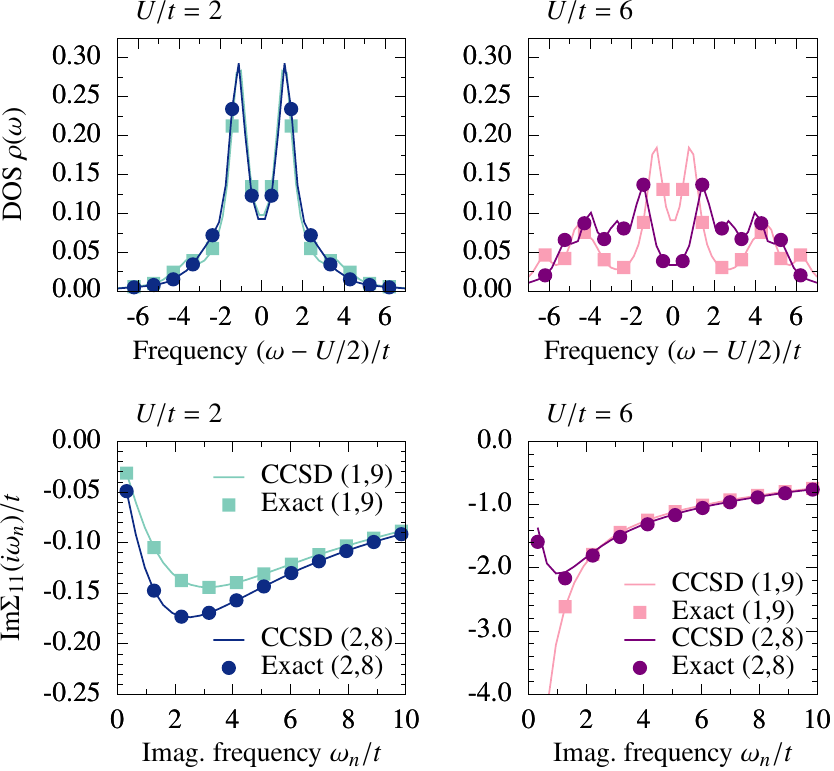}
\caption{
Single-site DMFT and two-site CDMFT results for the 1D Hubbard model,
comparing the use of the CC and exact diagonalization (ED) impurity solvers. 
Numbers in parentheses indicate the number of impurity sites and the
total number of bath sites.
Results are shown at half-filling with weak
interactions ($U/t=2$, left) and strong interactions ($U/t=6$, right).
The chemical potential is fixed at $\mu=U/2$ and the DOS is plotted with
a broadening of $\eta=0.5t$ for clarity.
}
\label{fig:hubbard_1d_exact}
\end{figure}

We first assess the accuracy of CCSD compared to exact diagonalization.  In
Fig.~\ref{fig:hubbard_1d_exact}, we show the DOS and self-energy from single-site DMFT with 9
bath sites and from two-site CDMFT with 4 bath sites per impurity site (8 in
total). Both correspond to an impurity problem with 10 sites which
is readily accessible with exact
diagonalization. At both values of $U/t$, the RCCSD and ED plots
are indistinguishable. Note that at small $U/t$, UHF does not break spin symmetry and UCCSD and RCCSD give
identical results.  At large $U/t$, UHF strongly breaks spin symmetry,
however the mean-field AFM order is reduced by UCCSD such that the final
results are almost indistinguishable from those of RCCSD.
Therefore, all results for the 1D Hubbard model are presented for RCCSD only.

\begin{figure}[t]
\centering
\includegraphics[scale=1.0]{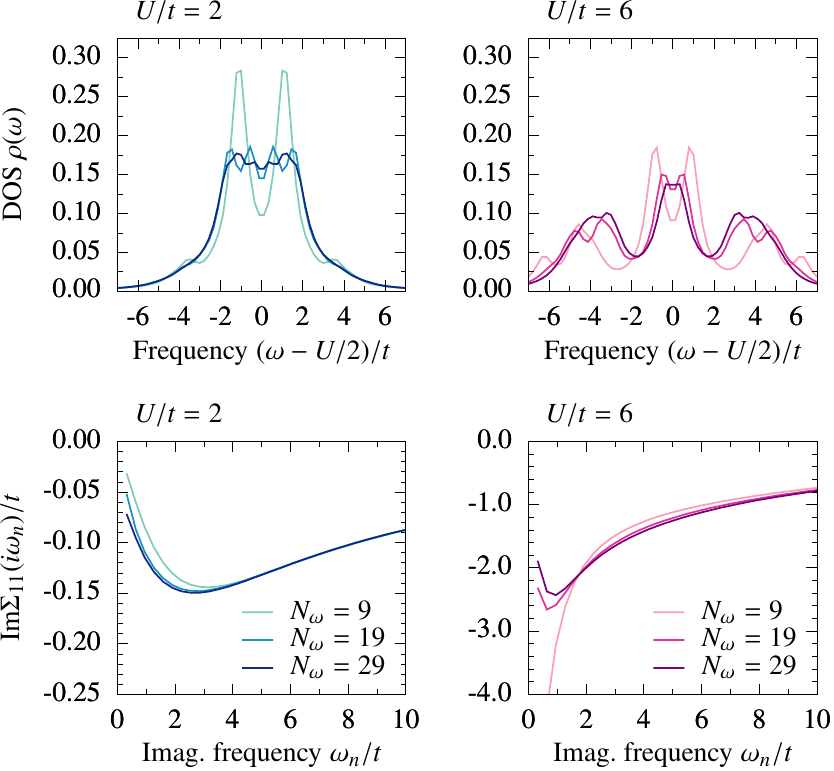}
\caption{
Single-site DMFT results for the 1D Hubbard at half-filling with weak
interactions ($U/t=2$, left) and strong interactions ($U/t=6$, right),
showing convergence with respect to the number of bath sites $N_\omega$.
The chemical potential is fixed at $\mu=U/2$ and the DOS is plotted with
a broadening of $\eta/t=0.5$ for clarity.
}
\label{fig:hubbard_1d_nimp-1}
\end{figure}

We next assess convergence of the DOS with respect to the number of bath sites,
which requires impurity problem sizes beyond the reach of ED.
In Fig.~\ref{fig:hubbard_1d_nimp-1}, we show the single-site DMFT DOS and self-energy computed using
9, 19, and 29 bath sites and the RCCSD solver. The largest impurity problem involves 30 particles in 30 orbitals.
The plots are qualitatively converged with 19 bath sites and 
converged to the eye with 29 bath sites.
Consistent with previous studies~\cite{Rozenberg1996,Bolech2003,Zgid2012}, we find that the single-site
DMFT produces a Kondo-like resonance in the DOS at large $U$, as compared with 
exact Mott insulating behavior at all $U$. We note that this Kondo-like behavior is a result of using a single impurity site in DMFT and is not due to the CCSD approximation.

\begin{figure}[t]
\centering
\includegraphics[scale=1.0]{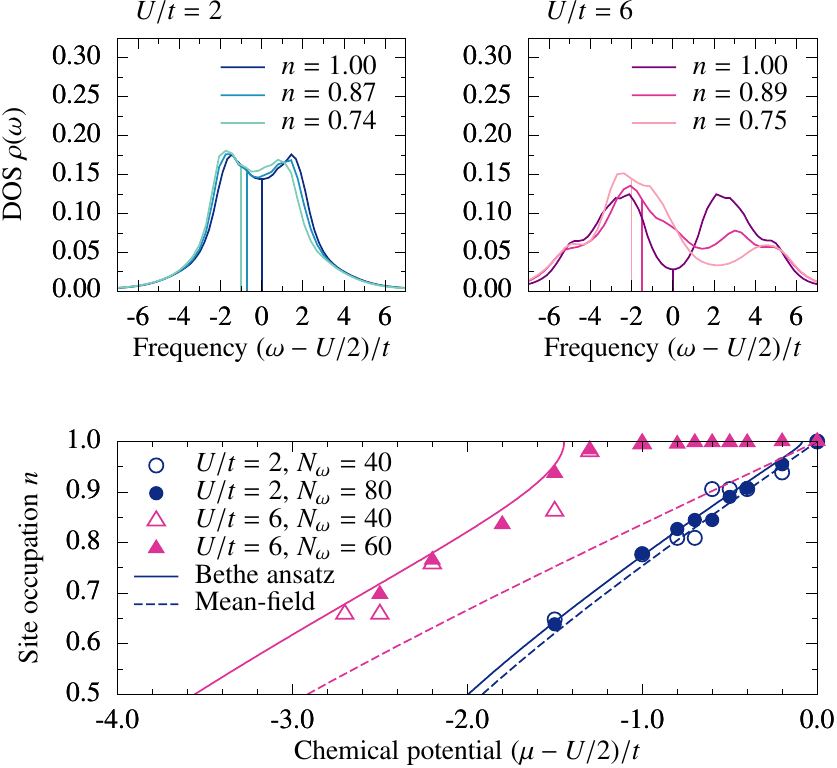}
\caption{
Two-site CDMFT results as a function of doping.  The DOS obtained using
$N_\omega=30$ (60 bath sites in total) is shown in the top two panels, where the
chemical potential is indicated by a vertical line.  The site occupancy as a
function of the chemical potential using $N_\omega=40$ (open symbols) and
$N_\omega=60,80$ (filled symbols) is shown in the bottom panel, along with the
numerically exact result from Bethe ansatz~\cite{Lieb1968} and the mean-field
result.
}
\label{fig:hubbard_1d_nimp-2_n}
\end{figure}

Based on the converged results for single-site DMFT, we next use
two-site CDMFT with at least 30 bath sites per impurity, i.e.~60 bath sites in total.
In Fig.~\ref{fig:hubbard_1d_nimp-2_n}, we present the DOS as a function
of occupancy.
At half-filling ($n=1$, $\mu=U/2$), we see a clear Mott gap proportional to 
$U/t$.  We access other filling fractions by changing the chemical
potential.  At small $U/t$, only minor changes are seen in the DOS for
moderate changes in the occupancy.
At larger $U/t$, there is significant redistribution of the spectral
weight towards lower energy, creating a metallic DOS around the
chemical potential, which is indicated by a vertical line in the DOS plots.
In the bottom panel of Fig.~\ref{fig:hubbard_1d_nimp-2_n}, we show the
occupancy as a function of the chemical potential for the two values
of $U$ studied.  
The discrete nature of the bath allows access to only a discrete set
of occupations, and so this latter data was obtained using larger
bath sizes.
The appearance of a Mott plateau and suppressed
compressibility is clearly seen for $U/t=6$. 
Due to the discretized bath, there are artificial plateaus in the regions where the Bethe ansatz (BA) varies monotonically, even with 40 bath sites per impurity site. By increasing the bath size to 80 and 60 bath sites per impurity site for $U/t=2$ and $6$, we show that such artificial behavior can be removed, and the occupancy converges to the BA solution. This result may suggest that when discretizing the hybridization along real axis, static quantities (e.g. occupancy) converge more slowly with respect to the number of bath orbitals as compared to dynamic quantities
  (such as the spectral function) although a more careful comparison with imaginary axis techniques is required.

\subsection{2D Hubbard model}

\begin{figure}[t]
\centering
\includegraphics[scale=1.0]{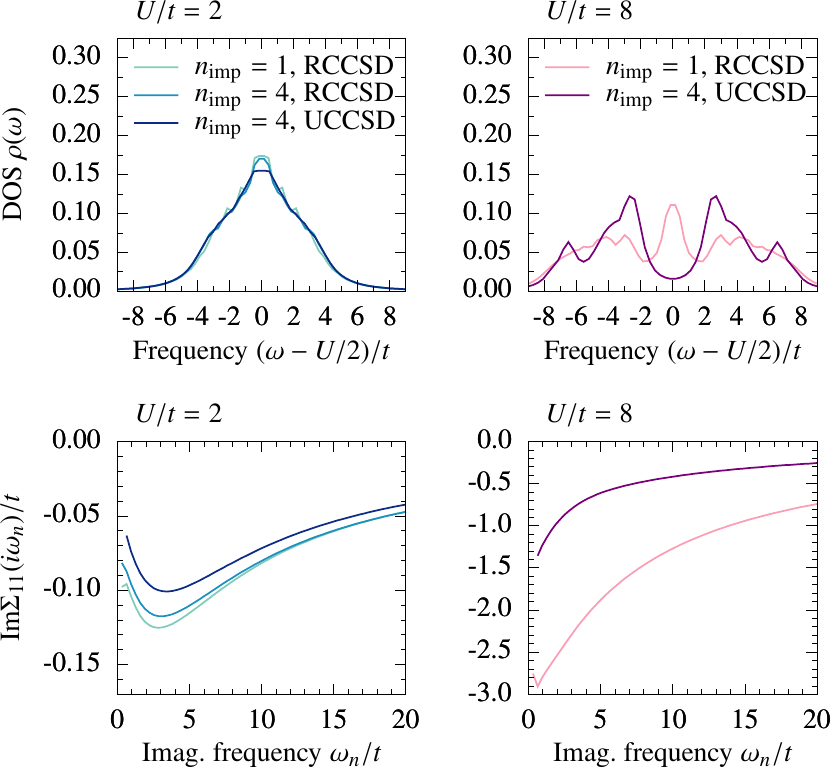}
\caption{
Single-site DMFT and four-site ($2\times 2$) CDMFT results for the 2D Hubbard at
half-filling with weak interactions ($U/t=2$, left) and strong interactions
($U/t=8$, right). We used 30 bath sites for single-site DMFT and 120 bath sites for the four-site CDMFT.
The chemical potential is fixed at $\mu=U/2$ and the DOS is
plotted with a broadening of $\eta/t=0.5$ for clarity.
}
\label{fig:hubbard_2d}
\end{figure}

We next study the 2D Hubbard model at half-filling.  We perform
calculations at $U/t=2$ and $U/t=8$. This is above and below the $2\times 2$ CDMFT 
paramagnetic Mott transition around $U/t\approx 6$~\cite{Park2008,Sordi2010}.  In
Fig.~\ref{fig:hubbard_2d}, we show results for single-site DMFT and
$2\times 2$ CDMFT. As before, we fix the number of bath sites
to be about 30 per impurity site, such that our largest calculations have 120 particles
in 120 orbitals. 

At $U/t=2$ DMFT and CDMFT give very similar DOS and self-energies; i.e.
the effect of the impurity size is small. Even if we allow for symmetry breaking
in the $2\times 2$ cluster, we find it to be very weak, thus (paramagnetic) RCCSD and (weakly antiferromagnetic) UCCSD give very similar
results. In both cases, the gap is smaller
than the bath discretization and thus indistinguishable from a metal.
At $U/t=8$, paramagnetic single-site DMFT with RCCSD or UCCSD gives identical
results, showing a Kondo-like feature centered between the lower and upper Hubbard bands.
Paramagnetic  2$\times2$ CDMFT with RCCSD fails to converge
in the DMFT loop.  The first few iterations of the DMFT cycle are physically reasonable
and show the opening of a paramagnetic Mott gap.  However, eventually an impurity problem is constructed
for which the iterative solution of the ground-state RCCSD equations fails to converge, because the associated CC amplitudes
become large. 
If we instead carry out an antiferromagnetic CDMFT calculation allowing for symmetry breaking in
the UCCSD solver, the DMFT cycle converges smoothly. As expected, the antiferromagnetic  DOS exhibits a gap, with a qualitatively
reasonable size of about $6t$. 

\section{Conclusions}

We have demonstrated the promise of coupled-cluster (CC) theory
as an impurity solver in single-site dynamical mean-field theory and multi-site cluster dynamical mean-field theory.
In particular, the polynomial cost of truncated CC allows the use of many
bath sites per impurity site, which enables a faithful discretization
of the hybridization directly on the real frequency axis.  In our studies of the Hubbard model,
we find that the Green's functions and self-energies are well converged using approximately 30 bath sites per impurity site.  

Despite the strongly correlated nature of the lattice problem, the low-order CCSD 
truncation provides very accurate results for spectral functions, self-energies, 
and occupation numbers as an impurity solver in DMFT, when there is a modest number of impurities. This is consistent with
the impurity problem being less strongly correlated than the lattice problem, even for a dense discretization of the hybridization.  
Consequently, we expect that CC impurity solvers will find most use with clusters of
a moderate size in real-space.  The low cost
of the CC impurity solvers also makes them very promising for
applications to real materials with many electrons and many orbitals per impurity site
in ab initio quantum chemical formulations of DMFT~\cite{Zgid2011}.
In particular, the embedding approach to
CC response functions in solids is a promising
alternative to full periodic CC calculations~\cite{McClain2017,Gruber2018},
and work along these lines is currently in progress.

\begin{acknowledgments}
  TZ and GKC were supported by the US Department of Energy, Office of Science, via grant number SC19390.
  GKC is also supported by the Simons Foundation, via the Many-Electron Collaboration, and via the Simons Investigator Program.
The Flatiron Institute is a division of the Simons Foundation.
\end{acknowledgments}

\appendix*
\renewcommand\thefigure{A.\arabic{figure}}
\setcounter{figure}{0}

\section{Causality in coupled-cluster theory}

\begin{figure}[htb!]
\centering
\includegraphics[scale=1.0]{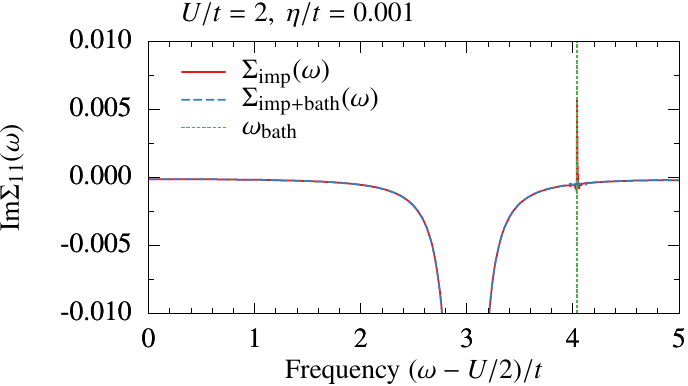}
\caption{
Impurity self-energy on the real frequency axis from a two-site CDMFT
calculation for the 1D Hubbard at half-filling with $U/t=2$. RCCSD is used as
the impurity solver and two bath sites are coupled to each impurity site. A
small broadening of $\eta=0.001t$ is used to show the non-causal behavior.
}
\label{fig:causality}
\end{figure}

As discussed in Sec.~\ref{sec:cc}, EOM-CCSD is not a conserving
approximation and thus the resulting Green's functions and self-energies need not
be causal.  The lack of strict causality is not seen in any of the results we
showed above. However, it is possible to observe small violations of causality
if we carry out calculations with a very small broadening (e.g. $\eta=0.001t$).
An example of non-causal behaviour that can occur in this
setting is shown in Fig.~\ref{fig:causality}.  As defined in
Eq.~\eqref{eq:sigmaimp}, $\mathbf{\Sigma}_\mathrm{imp}(\omega)$ should always
have a negative imaginary part if it is computed from a causal impurity Green's
function~\cite{Biroli2004}. However, we see that the CCSD impurity self-energy
develops a positive imaginary part exactly around the bath frequencies. One way
to understand this is that truncated CC theory does not include all
possible diagrams involving interactions and hopping to the bath (i.e. the
hybridization). Thus the hybridization contribution to
$\mathbf{G}_\text{imp}^{-1}$ in Eq.~\eqref{eq:sigmaimp} and
$\mathbf{\Delta}(\omega)$ do not precisely cancel to give a causal self-energy.

This non-causality only appears when the broadening is very small and does not
affect the results we presented.  Nevertheless, we observe that there is a
simple procedure that removes this issue in practice. Instead of computing
$\mathbf{\Sigma}_\mathrm{imp}(\omega)$  in Eq.~\eqref{eq:sigmaimp}, one can
first compute the self-energy for the whole impurity plus bath system
\begin{align}
\mathbf{\Sigma}_{\imp+\bath}(\omega) 
    = \mathbf{G}_{0, \imp+\bath}^{-1}(\omega) -\mathbf{G}_{\imp+\bath}^{-1}(\omega).
\label{eq:sigmaib}
\end{align}
where $\mathbf{G}_{0, \imp+\bath}^{-1}(\omega)$ is the non-interacting Green's
function of the impurity plus bath system.  The impurity self-energy is then
defined as the impurity block of $\mathbf{\Sigma}_{\imp+\bath}(\omega)$. (In an
exact solver, this is the only non-zero part of
$\mathbf{\Sigma}_{\imp+\bath}(\omega)$). As shown in Fig.~\ref{fig:causality},
the non-causal behavior disappears if the impurity self-energy is defined in
this way, at the cost of more computation. While we do not claim that this
procedure makes the self-energy strictly causal under all circumstances, we find
that it removes even the small degree of non-causal behaviour observed at very
small broadenings in this work.

%

\raggedbottom

\end{document}